\documentclass[fleqn,12pt,twoside]{article}
\usepackage{espcrc1}

\usepackage{graphicx}
\usepackage[figuresright]{rotating}

\newcommand{\AmS}{{\protect\the\textfont2
  A\kern-.1667em\lower.5ex\hbox{M}\kern-.125emS}}

  \newcommand \beq{\begin{eqnarray}}
\newcommand \eeq{\end{eqnarray}}
\newcommand{\bea}{\begin{eqnarray}}
\newcommand{\eea}{\end{eqnarray}}

\def\simle{\mathrel{\rlap{\raise 0.511ex \hbox{$<$}}{\lower 0.511ex \hbox{$\sim$}}}}
\def\simge{\mathrel{ \rlap{\raise 0.511ex \hbox{$>$}}{\lower 0.511ex \hbox{$\sim$}}}}

\def\tr{{\,\rm tr\,}}
\def\0{\over } \def\2{{1\over2}} \def\4{{1\over4}}
\def\5{\hat } \def\6{\partial }

\def\8#1{{\textstyle{#1}}}

\def\({\left(} \def\){\right)} \def\<{\langle } \def\>{\rangle }

\title{High temperature phase of QCD}

\author{Jean-Paul Blaizot \address{ECT*, Villa Tambosi, \\
        strada delle Tabarelle, 286, I 38050 Villazzano (TN), Italy}%
        \thanks{Member of CNRS, France},
        }

\begin{document}

% typeset front matter
\maketitle

\begin{abstract}
I give a brief overview of our present understanding of the  high temperature phase of QCD, trying to clarify some of the theoretical issues involved in the current discussions that emphasize the strongly coupled character of the quark-gluon plasma produced at RHIC.
\end{abstract}

\section{From the ``ideal gas'' to the ``perfect liquid''}

QCD asymptotic freedom leads us to expect that  matter at very high temperature and/or density should become simple \cite{Collins:1974ky,Cabibbo:1975ig}: an ideal gas of quarks and gluons whose free motion is only weakly perturbed by their interactions.  Lattice calculations  indeed  show that, at very high temperature, the thermodynamical functions indeed go, albeit slowly, towards those of free massless particles (see for instance \cite{Karsch:2001cy}). We also start to have a semi-analytic understanding of this high temperature state  in terms of quasiparticles, as we shall discuss later.
In agreement with general theoretical considerations on the running of the effective coupling with the temperature, lattice calculations also show that as the temperature decreases, the coupling increases  and the physics becomes less and less perturbative. Eventually at some critical temperature, a phase transition, or a rapid cross-over, takes place towards a regime dominated by hadronic degrees of freedom.

This standard, and theoretically well established, picture has been recently challenged, with most of the present discussions emphasizing the  strongly coupled character of the quark-gluon plasma.  Three elements have conspired to this shift in paradigm. First, the RHIC data do not provide any evidence for ideal gas behavior, but are better interpreted by assuming that the produced matter behaves as a liquid. Second, perturbation theory is notoriously unable to describe the quark-gluon plasma unless the temperature is extremely high. Third, new techniques have emerged that allow strong coupling calculations to be done in  non abelian gauge theories. 

Most proposed ``signatures"  that have been suggested during the preparation of the heavy ion program, are  based on  the simple  picture of  the quark-gluon plasma as an ideal gas of weakly interacting constituents. Part of the reason for this is simple: this is a regime where one can do first principle (and elementary) calculations. However, RHIC is forcing us to look into a region where theory is hard, i.e., study the quark-gluon plasma in a regime where the coupling is not very small. The temperatures reached at RHIC are presumably not high enough to reach the asymptotically free regime. 
We learned from RHIC (see \cite{Arsene:2004fa,Back:2004je,Adcox:2004mh,Adams:2005dq}, and also  \cite{Gyulassy:2004zy,Shuryak:2004cy}) that the matter produced in ultrarelativistic heavy ion reactions has a large energy density, high enough for  the early stages of the collisions to be dominated  by partonic degrees of freedom. Suppression of jets reveals also strong densities and strong interactions. The observed elliptic flow indicates that the matter behaves collectively, as a liquid with low viscosity \cite{Shuryak:2003xe}.  Although it is difficult to place quantitative bounds on the values of the viscosity from the analysis of the data, it seems indeed that low values of the viscosity are needed \cite{Teaney:2003kp}, values that  seem to be incompatible with perturbative QCD calculations.

 Much effort has been put into calculating the successive orders of the perturbative expansion for the pressure \cite{Arnold:1995eb,Zhai:1995ac,Braaten:1996jr,Kajantie:2002wa} and the series is known now up to order $g^6\ln g$\cite{Kajantie:2002wa}. Such calculations indicate that the asymptotic quark-gluon plasma indeed emerges, as expected, at very high temperature. However, they also show that perturbation theory  makes sense only for very small couplings, corresponding to extremely large values of $T$. For not too small couplings, the successive terms in the expansion oscillate wildly and the dependence of the results on the renormalization scale keeps increasing order after order \cite{Blaizot:2003tw}. This situation is to be contrasted with what happens at zero temperature, where perturbative calculations achieve a reasonable accuracy already at the GeV scale. At finite temperature,  thermal fluctuations  alter the infrared behavior in a profound way. Nevertheless, as I shall argue, this does not imply that weak coupling calculations are useless. In fact a lot can be learned from them, in particular how to capture the right physics that can allow for smooth extrapolations to the strong coupling regime. 

Finally, new techniques for doing calculations in strong coupling have emerged recently, giving hope that we may soon have at our disposal new tools to handle strongly coupled gauge theories. These techniques are based on the  
recognition that some supersymmetric Yang-Mills theories are dual to 
gravitation theories. The duality involves an interchange of the
regimes of weak and strong coupling: weak coupling in the
gravity theory corresponds to strong coupling in the gauge
theory. This duality offers the possibility to study strongly coupled gauge theories by performing perturbative calculations in their gravity duals.  This possibility has been exploited in a number
of recent publications. One  prediction of such calculation is
that the entropy $S$ behaves, in strong coupling as \cite{Gubser:1998nz}
\beq
\frac{S}{S_0}=\frac{3}{4}+\frac{45}{32}\zeta(3)\frac{1}{\lambda^{3/2}},
\eeq where $\lambda\equiv 2g^2 N_c$ and $S_0$ is the entropy of the non interacting system . Thus, 
 in the limit of strong coupling, $\lambda\to\infty$,
the entropy is bounded from below by the value $S/S_0=3/4$. The fact that this value is close to that
obtained in lattice calculations at temperature above $T_c$ has
contributed to sustain the  speculations that the quark-gluon  plasma above $T_c$ is in  a
strongly coupled regime.  Note however that at $T\simeq 3T_c$, the value of the entropy density is half-way between its weak coupling value  and its strong coupling value (see Fig.~\ref{figSg} below), so there is no compelling reason to favor at that point an approach based on a strong coupling expansion. Furthermore, the  minimal value of $3/4$ is obviously not compatible
with lattice data near $T_c$ where the entropy vanishes and the coupling is the strongest. In fact, supersymmetric Yang-Mills theories have
symmetries that make them rather different from QCD: in particular the coupling
constant does not run, and there is no phase transition. The running of the coupling constant is however an essential property of QCD; in particular, above the transition region  it is accompanied by a breaking of conformal symmetry that is 
very well observed in lattice calculations
\cite{Boyd:1996bx,Gavai:2004se,Gavai:2005da}.

In fact the physics above $T_c$, in the temperature range $T_c\simle T\simle 3T_c$, remains poorly understood, with many unanswered questions:  What
are the degrees of freedom relevant for an effective theory? Are remnants of confinement important, such as those captured by  effective theories for the  Polyakov loop  (see for instance \cite{Pisarski:2000eq,Vuorinen:2006nz} or, on a more phenomenological side, \cite{Ratti:2005jh})?
What is
the role, if any, of  bound states \cite{Shuryak:2004tx}? What is their fate as the temperature grows, do they survive at large temperature, as recent lattice data
suggest \cite{Asakawa:2003re}? What are the charge carriers \cite{Ejiri:2005wq,Gavai:2005yk}? Etc.

In the rest of this talk, I shall not discuss this transition region, but focus on the high temperature phase. Also I shall only consider the regime of vanishing chemical potential (for a recent study of QCD thermodynamics at finite chemical potential, see \cite{Ipp:2006ij}).

\section{Weakly or strongly coupled quark-gluon plasmas}

In order to get orientation into the effects of the interactions a general strategy  is to compare the kinetic energy of the particles to their interaction energy. When the kinetic energy dominates, the non-interacting system constitutes a good starting point for the theoretical description, and one may think of using perturbation theory. Note that this kind of argument may be obscured by the emergence of new degrees of freedom in the interacting system: in this case one could be led to conclude that it is strongly coupled, while the main effect of the interactions is to generate these new degrees of freedom which may have weak residual interactions. The natural starting point of the theoretical description in such cases would be the  non-interacting system  built with  these new degrees of freedom.

Consider an elementary plasma, made of electrons,  positrons and photons.
When the charged particles behave individually as
classical particles, the mean kinetic energy  of an electron is simply proportional to the temperature $T$, while the average potential energy per particle  is of the order $e^2 n^{1/3}$, where $n$ is the number density, so that $n^{1/3}$ is the inverse of the interparticle distance. The condition that the plasma be ideal is then $T\gg e^2 n^{1/3}$. The dimensionless parameter 
 $g^2\equiv e^2 n^{1/3}/T$ is essentially the plasma parameter $\Gamma$ used in plasma physics \cite{Ichimaru1982}. In terms of $g$, the condition that the plasma is ideal is simply $ g\ll 1$.
The Debye screening length is
$
\lambda\sim\sqrt{{T}/{ne^2}}\sim n^{-1/3}/g.
$
Thus when $g$ is small, the screening length is large compared to the interparticle distance, which is a criterion for collective behaviour. Also when $g$ is small individual collisions can be ignored.

In ultrarelativistic plasmas, the temperature and the density are no longer independent control parameters since   $n\sim T^3$. Then the plasma is characterized by a single dimension-full parameter, $T$. This is the situation in QCD at high temperature,  where  the parameter $g$ is  the gauge coupling (at a scale $\sim T$). Note that, just above $T_c$, the couping constant is not small, but not huge either, $g\simeq 2$ (see e.g. \cite{Laine:2005ai}). However, to decide whether the quark-gluon plasma is strongly or weakly coupled it  is  essential to recognize that  the effects of the interactions depend not only  on the strength of the coupling but also on the magnitude of the thermal fluctuations, which depends on their wavelengths.  At weak coupling a hierarchy of scales of thermal fluctuations emerges, each scale being associated with well identified physics \cite{Blaizot:2001nr}. Predicting the effect of the interactions amounts to comparing the kinetic energy $\sim \langle(\partial A)^2\rangle\sim k^2\langle A^2\rangle$ with interaction energy $g^2\langle A^4\rangle\sim g^2 \langle A^2\rangle^2$, or equivalently $k^2$ with $g^2\langle A^2\rangle$.

The  plasma particles  have typical energy and momentum of the order of the temperature  $k\sim T$. The thermal fluctuations at this scale are  $\langle
A^2\rangle_T\sim T^2$. These  {\it hard} fluctuations constitute the  dominant contribution to the  energy density, but they produce only a small perturbation on the motion of a plasma particles if $g$ is small. 
Consider now a {\it soft}  excitation at the momentum scale  $k\sim gT$. The fluctuations at this scale are $\langle A^2\rangle_{gT} \sim
gT^2$. The self-interactions of soft modes remain a small correction ($g^2 \langle A^2\rangle_{gT} \sim g^3 T^2 \ll k^2\sim g^2 T^2$), but the interaction of the soft mode with the hard fluctuations cannot be ignored  ($g^2 \langle A^2\rangle_{T} \sim g^2 T^2 \sim k^2 $): the hard particles renormalize non-perturbatively the propagation of the soft modes; this is the physical origin of the hard thermal loops. And indeed, the scale
$gT$ is that  at which collective phenomena develop. The emergence of the Debye
screening mass
$m_D\sim gT$ is the simplest example  of such phenomena.
Moving down to a lower momentum scale, one meets the contribution of the
unscreened magnetic fluctuations which play a dominant role for
$k\sim g^2T$.  At that scale, 
$\langle A^2\rangle_{g^2T}
\sim\,g^2 T^2,$
so that $g^2 \langle A^2\rangle_{g^2T}\sim g^4T^2\sim k^2$: the fluctuations are no longer perturbative. This
is the origin of the breakdown of perturbation theory that occurs at order $g^6$ in the pressure ($\sim g^2\langle A^2\rangle_{g^2T}^2$).

This particular pattern of scales, and the way they are coupled,  is what complicates the theoretical description of the quark-gluon plasma. Even when the coupling is weak, non-perturbative effects arise: for instance collective modes develop, whose treatment   requires  resummations or the use of effective theories. The real issue here is not so much the strength of the coupling but the fact that there are many quasi degenerate active degrees of freedom. As we shall see in the rest of the talk, weak coupling techniques allow us to identify and perform  the appropriate reorganizations of the perturbative expansion that are needed  in order to take into account the role of the  new degrees of freedom, in particular the collective modes. We shall  see that once this is done, the extrapolation to strong coupling is much smoother than what perturbation theory could lead us to expect. 

\section{Weak coupling techniques for strong coupling regimes, resummations}

Since the major problems in thermal QCD come from the infrared sector, and in particular from modes carrying zero Matsubara frequencies, it is natural to isolate those modes and construct for them an effective theory through a  procedure known as dimensional reduction. The resulting Lagrangian is of the form \cite{Appelquist:1981vg,Nadkarni:1983kb,Nadkarni:1988fh}
\beq\label{LQCDdr} \mathcal L_E=\2 \tr F_{ij}^2 + \tr [D_i,A_0]^2
+m_E^2 \tr A_0^2 + \2 \lambda_E (\tr A_0^2)^2 +\ldots \eeq
where the parameters $g_E$ ($D_j=\partial_j+ ig_E A_j$), $m_E$ and $\lambda_E$ are determined perturbatively as a function of the gauge coupling $g$ by matching. In lowest order  
$g_E^2 = g^2T$ and $ m_E^2\sim g^2 T^2$, 
$\lambda_E\sim g^4T$.  

Calculations based on this scheme have been pushed to high order \cite{Kajantie:2000iz},
but they depend on an as yet undetermined 4-loop matching
coefficient. By adding a parameter to account for the uncalculated $g^6$ contribution, one can match the four-dimensional lattice results. The required value of the coefficient is not very large, suggesting that the contribution of the magnetic sector to the pressure is numerically not important at high temperature. 

An interesting feature of the perturbative calculation when
organized through the dimensionally reduced effective theory
(\ref{LQCDdr}) is that the large scale dependences of strict
perturbation theory can be reduced when the effective parameters are
not subsequently expanded out\cite{Kajantie:2002wa,Blaizot:2003iq}, with noticeable improvements when going from two-loop to three-loop in the effective theory.

Other ways to reorganize the perturbative expansion have been tried.
One proposal, called ``screened
perturbation theory'' \cite{Karsch:1997gj,Andersen:2000yj},  has
been generalized to QCD by Andersen, Braaten, and Strickland
\cite{Andersen:1999fw,Andersen:2002ey}. A more systematic approach to screened perturbation theory is based on  an expression for the
entropy density that can be obtained from a $\Phi$-derivable
two-loop approximation \cite{Blaizot:1999ip,Blaizot:2000fc}. This approach has a clear physical content: the dominant effect of the interactions is to turn  the original degrees of freedom, quarks and gluons, into massive quasiparticles, with weak residual interactions.  This theoretical approach gives support to the more phenomenological one proposed in Ref. ~\cite{Peshier:1995ty}

\begin{figure}[tb]
\centerline{\includegraphics[bb=70 200 540
560,width=7.5cm]{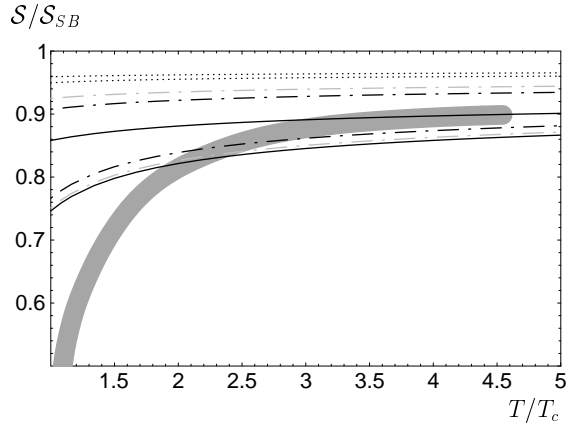}} \caption{Comparison of the lattice data
for the entropy of pure-glue SU(3) gauge theory of
Ref.~\protect\cite{Boyd:1996bx} (gray band) with the range of ${\cal
S}_{HTL}$ (solid lines) and ${\cal S}_{NLA}$ (dash-dotted lines) (see text). From Ref.~\cite{Blaizot:2000fc}.}
\label{figSg}
\end{figure}

In
Refs.~\cite{Blaizot:1999ip,Blaizot:2000fc} the lattice results for
the thermodynamic potential for $T \ge 3 T_c$ were quite well
reproduced as can be seen in  
Fig.~\ref{figSg}. In this figure,  numerical results are given for the  entropy obtained in two successive approximations: in the first, denoted HTL, the hard thermal loop self-energies are used as an input of the calculation, in the second, denoted as  NLA, corrections to the HTL self-energies are taken into account.  The full lines show the
range of results for ${\cal S}_{HTL}$ when the renormalization scale
$\bar\mu$ is varied from $\pi T$ to $4\pi T$; the dash-dotted lines
mark the corresponding results for ${\cal S}_{NLA}$. The dark-gray
band are lattice data from Ref.~\cite{Boyd:1996bx} (the more recent
results from Ref.~\cite{Okamoto:1999hi} are consistent with the
former within error bars). Evidently, there is very good agreement for $T\simge 3T_c$. The result from Ref.~\cite{Andersen:1999fw,Andersen:2002ey} is indicated by the dooted lines at the top of Fig.~\ref{figSg}.

The 2PI formalism used in the calculation of the entropy that we have just described  has been tested \cite{Blaizot:2005wr} in a model with a large number of quark flavors, which can be solved exactly. In this case, the 2PI formalism that we used is exact. The comparison provides then indications on the quality of the further approximations involved in solving the 2PI equations. The results are quite encouraging, and a remarkable agreement can be achieved for quite large values of the coupling constant. Note however, that these calculations indicate that much accuracy can be gained by taking fully into account the momentum dependence of the corrections to the asymptotic thermal masses that enter the calculation of the NLA (in the results reported in Fig.~\ref{figSg}, only momentum averages have been used  \cite{Blaizot:1999ip,Blaizot:2000fc}). Implementing such corrections is technically demanding and has not been completed yet.

\section{Insights from the functional renormalization group}

In the last part of this talk, I would like to examine the issues addressed above from the perspective of  the functional renormalization group (fRG)  \cite{Wetterich:1992yh,Ellwanger:1993kk,Tetradis:1993ts,Morris:1993qb,Morris:1994ie} (sometimes called exact or non-perturbative, depending on which aspects of the formalism one wants to emphasize).

The fRG  has been applied, in various
incarnations, to a variety of  problems (for reviews see
\cite{Bagnuls:2000ae,Berges:2000ew}). It allows for non-perturbative
approximation schemes which can, in some instances, be quite accurate. Functional RG
techniques have been applied to problems at finite temperature in the past
\cite{Tetradis:1993ts,Reuter:1993rm,Andersen:1999dy,D'Attanasio:1996fy,Liao:1995gt},
and the general behaviors that we shall discuss here have
been known already for some time. However, the main focus of
previous studies has been the description of the phase transitions,
rather than the specific problem that we  address here.

\begin{figure}[htb]
\begin{center}\leavevmode
\includegraphics[width=7cm]{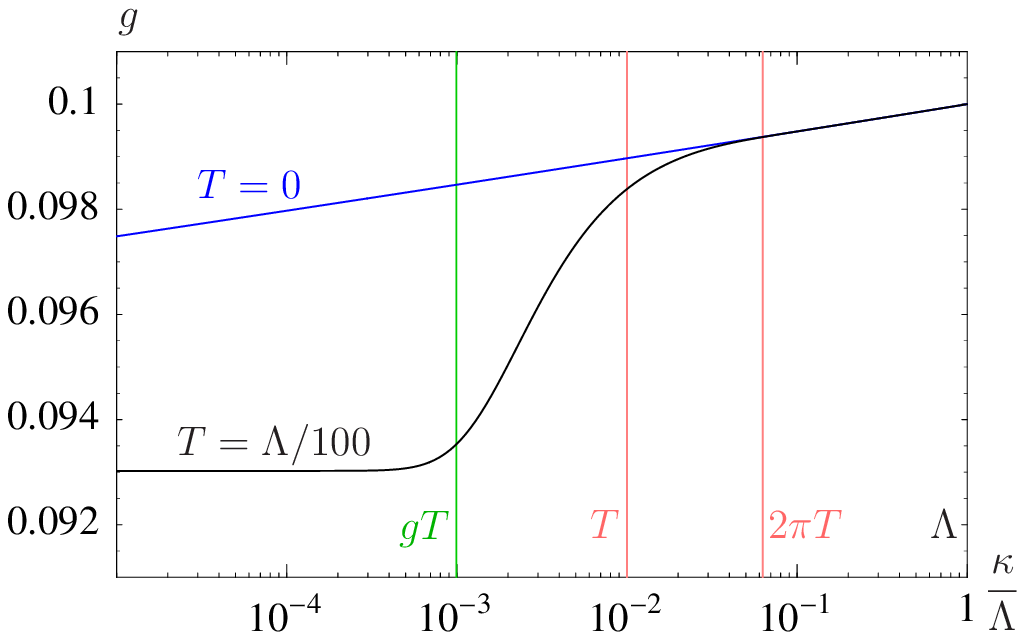}
\includegraphics[width=7cm]{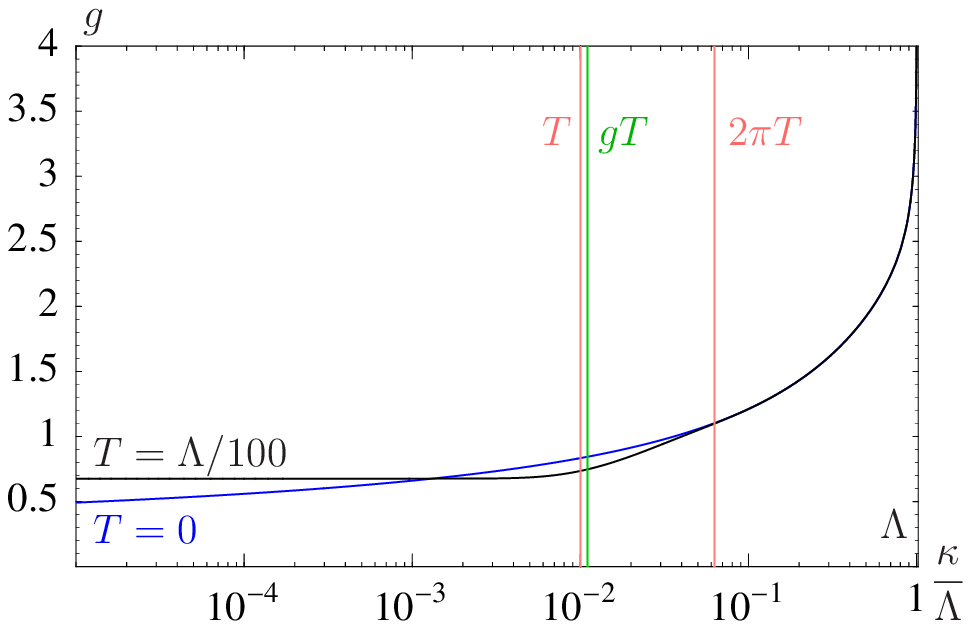}
\end{center}
\caption{The flow of the coupling constant as the function of the ratio $\kappa/\Lambda$, where $\kappa$ is the scale of the infrared cut-off, while $\Lambda$ is the microscopic  scale. The left figure corresponds to weak coupling ($g(2\pi T)= 0.0994$), while the right figure corresponds to strong coupling ($g(2\pi T)=1.1$).\label{flowcoupling}}
\end{figure}

There is some analogy between the effective field theory approach
and the functional renormalization group: in effective field theory
one integrates out degrees of freedom above some cut-off; in the
renormalization group this integration is done smoothly.  The general strategy of the fRG  is to control the long wavelength fluctuations with an infrared regulator depending on a continuous parameter $\kappa$. When $\kappa$ is of the order of the microscopic scale, fluctuations are essentially suppressed. On the other hand when $\kappa\to 0$ they are all included. The presence of the regulator allows us to write flow equations that can be integrated all the way down to $\kappa=0$ in order to get the physical quantities. 
 In a way,
the renormalization group builds up a continuous tower of effective
theories that lie infinitesimally close to each other and are labeled by
the momentum cut-off scale $\kappa$. These effective theories are
related by a renormalization group flow equation.  This picture is independent of the value of the coupling, and as we shall see,   the
renormalization group provides  a smooth extrapolation from the regime
of weak coupling, characterized by a clean separation of scales, towards the strong
coupling regime where all scales get mixed.

 I shall only present a few results obtained recently together with A. Ipp, R. Mendez-Galain, and N. Wschebor \cite{BIMW}. These results  concern   a scalar  $\phi^{4}$ theory   with $O(N)$
symmetry which exhibits similar bad convergence properties of its thermal perturbative expansion as QCD.  In our study, we have used the so-called  local potential approximation (LPA) \cite{Berges:2000ew}.
We have shown  that this approximation is compatible with perturbation theory up to, and including, order $g^3$, and that is provides a smooth extrapolation to strong coupling similar of that of a simple 2PI approximation. (Note that in scalar theories, the presence of the Landau pole puts a limit on the maximum value of the coupling for wich calculations are meaningful. Strong coupling here means $g\simge 1$.)

\begin{figure}[htb]
\begin{center}\leavevmode
\includegraphics[width=8cm]{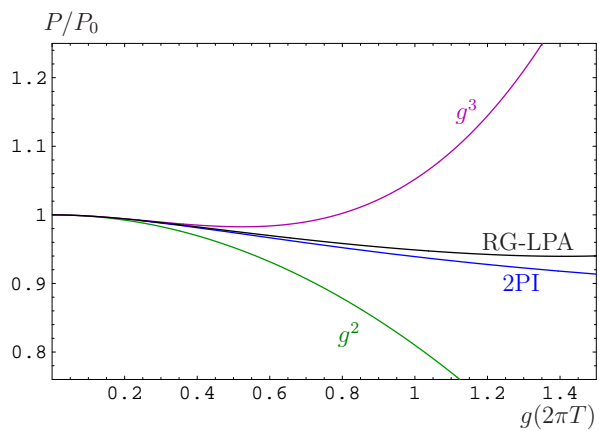}
\end{center}
\caption{Pressure as a function of the coupling $g(2\pi T)$. The result of the RG-LPA calculation is compared with the perturbative estimates of order $g^2$ and $g^3$ and also with a simple 2PI resummation. \label{pressure}}
\end{figure}

The left panel of Fig.~\ref{flowcoupling} illustrates the flow of the effective coupling as the momentum cut-off runs down to zero. One observes that as long as $\kappa\simge 2\pi T$, the running of the coupling is not modified by the thermal fluctuations. These fluctuations start to play a role when $\kappa\simle 2\pi T$, turning the flow into a  three dimensional one. This provides, in this context, a nice illustration of the phenomenon of dimensional reduction. When $\kappa$ becomes of order $gT$, the thermal mass provides an infrared cut-off and the flow stops (the infrared regulator playing then no role anymore). As revealed by the right panel of Fig.~\ref{flowcoupling}, when the coupling increases, the pattern does not change much: the thermal fluctuations start to play a role when $\kappa\simle 2\pi T$, turning the flow into a three dimensional one. The range of values of $\kappa$ where the flow is three dimensional is reduced because the thermal mass increases with the coupling. At the same time the amplitude of the three-dimensional flow is enhanced by the larger value of the coupling. These competing effects  contribute to the stability of the results as one goes from weak to strong coupling.

This is reflected in the plot of the pressure as  a function of the coupling, Fig.~\ref{pressure}. The pressure calculated with the fRG  is compared with the order $g^2$ and $g^3$ of perturbation theory, as well as   with the result of a simple self-consistent 2PI  approximation (this 2PI approximation, as well as the approximate solution of the flow equations become exact in the large $N$ limit).

%\bibliographystyle{unsrt}
%\bibliography{rhic,jpb,functionalRG,tft,tftpr,qft,mabiblio}

\end{document}